%% file: 2hp_aa.tex
\documentclass{aa}

\usepackage[english]{babel}
\usepackage{graphicx}
\usepackage[varg]{txfonts}
\usepackage{natbib}
\usepackage{longtable}
\usepackage{array}
\usepackage{multirow}
\usepackage{amsmath}

\bibpunct{(}{)}{;}{a}{}{,}
\begin{document}

\title{Millimetre-wave spectroscopy of 2-hydroxyprop-2-enal and an astronomical search with ALMA
\thanks{Table 5 is only available in electronic form at the CDS via anonymous ftp to cdsarc.u-strasbg.fr (130.79.128.5) or via http://cdsweb.u-strasbg.fr/cgi-bin/qcat?J/A+A/}
}

   \author{J.~Kouck\'{y}\inst{1}
   \and L.~Kolesnikov\'{a}\inst{1}
   \and K.~Lukov\'{a}\inst{1}
   \and K.~V\'{a}vra\inst{1}
   \and P.~Kania\inst{1}
   \and A.~Coutens\inst{2}
   \and J.-C.~Loison\inst{3}
   \and J. K.~Jørgensen\inst{4}
   \and A.~Belloche\inst{5}
   \and \v{S}.~Urban\inst{1}
          }

\institute{Department of Analytical Chemistry, University of Chemistry and Technology,
Technick\'{a} 5, 166 28 Prague 6, Czech Republic\\\email{jan.koucky@vscht.cz; lucie.kolesnikova@vscht.cz}
\and
Institut de Recherche en Astrophysique et Plan\'{e}tologie (IRAP), Universit\'{e} de Toulouse, UPS, CNRS, CNES, 9 av. du Colonel
Roche, 31028 Toulouse Cedex 4, France
\and
Institut des Sciences Mol\'{e}culaires (ISM), CNRS, Universit\'{e} Bordeaux, 351 cours de la Lib\'{e}ration, 33400 Talence, France
\and
Niels Bohr Institute, University of Copenhagen, Øster Voldgade 5–7, 1350 Copenhagen K, Denmark
\and
Max-Planck-Institut f\"{u}r Radioastronomie, Auf dem H\"{u}gel 69, 53121 Bonn, Germany}

\date{Received ; accepted }

\titlerunning{Millimetre-wave spectrum and ISM search of 2-hydroxyprop-2-enal}
\authorrunning{Kouck\'{y} et al.}


\abstract
   {Several sugar-like molecules have been found in the interstellar medium. The molecule studied in this work, 2-hydroxyprop-2-enal, is among the candidates to be searched for, as it is a dehydration product of C$_3$ sugars and contains structural motifs typical for some interstellar molecules. Furthermore, it has been recently predicted to be more abundant in the interstellar medium than its tentatively detected isomer 3-hydroxypropenal.}
   {So far, only low-frequency microwave data of 2-hydroxyprop-2-enal have been published.  The aim of this work is to deepen knowledge about the millimetre-wave spectrum of 2-hydroxyprop-2-enal in the region enabling its detailed search towards astronomical objects. In particular, we target the solar-type protostar IRAS16293–2422 and star-forming region Sagittarius~(Sgr)~B2(N).
      }
   {The rotational spectrum of 2-hydroxyprop-2-enal was measured and analysed in the frequency regions of 128--166 GHz and 285--329 GHz. The interstellar exploration towards IRAS16293-2422 was based on the Atacama Large Millimeter/submillimeter Array (ALMA) data of the Protostellar Interferometric Line Survey. We also used the imaging spectral line survey ReMoCA performed with ALMA toward Sgr~B2(N) to search for 2-hydroxyprop-2-enal in
   the interstellar medium. We modelled the astronomical spectra under the assumption of local thermodynamic equilibrium.}
   {We provide laboratory analysis of hundreds of rotational transitions of 2-hydroxyprop-2-enal in the ground state and the lowest lying excited vibrational state. We report its nondetection towards IRAS16293 B. The 2-hydroxyprop-2-enal/3-hydroxypropenal abundance ratio is estimated to be $\lesssim$ 0.9-1.3 in agreement with the predicted value of $\sim$ 1.4. We also report the nondetection of 2-hydroxyprop-2-enal toward the hot molecular core Sgr~B2(N1). We did not detect the related aldehydes 2-hydroxypropanal
   and 3-hydroxypropenal either. We find that these three molecules are at least 9, 4, and 10 times less abundant than acetaldehyde in this source, respectively.}
   {Despite the nondetections of 2-hydroxyprop-2-enal, the results of this work represent a significant improvement on previous investigations in the microwave region and meets the requirements for further searches of this molecule in the interstellar medium.}


   \maketitle
%

\section{Introduction}
\label{sect_intro}

The discovery of the first interstellar aldehyde, formaldehyde, dates back to 1969 \citep{Snyder1969}. Since then, eight more aldehydes, including their protonated, sulphur, and cyano analogues, have been detected by astronomical observations \citep{McGuire2022}. One of them is 2-hydroxyethanal \citep{Hollis2000a}, commonly known as glycolaldehyde, a prebiotic, sugar related molecule that fulfils the empirical formula C$_n$H$_{2n}$O$_n$ (or C$_n$(H$_2$O)$_n$), which initiated the search for sugars in the interstellar medium  (ISM). Glyceraldehyde, as the next member of the homologous series of carbohydrates, with the formula C$_3$H$_6$O$_3$, has not been successfully detected in the ISM so far \citep{Hollis2004,Jimenez-Serra2020a}. However, its presence there was suggested by laboratory experiments with interstellar ice analogues \citep{Layssac2020,Fedoseev2017}. The same molecular formula belongs to the other sugar molecule – dihydroxyacetone (glycerone, DHA), a simple ketotriose. It was identified in the Murchison meteorite \citep{Cooper2001} and therefore awakened interest in its interstellar search. The initial survey was performed using the Caltech Submillimeter Observatory with the tentative detection of nine emission lines in Sagittarius B2(N-LMH) \citep{WidicusWeaver2005}. However, other studies have not confirmed the presence of DHA there \citep{Jimenez-Serra2020a,Apponi2006}. The longer sugar with four carbon atoms, erythrulose, was another candidate to seek in the interstellar medium, regrettably with a negative outcome  \citep{Insausti2021}.

Lactaldehyde (2-hydroxypropanal, or hydroxypropionaldehyde), another oxygen bearing member of the C$_3$ family of molecules and a methyl derivative of glycolaldehyde, has been searched so far unsuccessfully in three star-forming regions \citep{Alonso19}.  Nevertheless, several attempts have been undertaken to discover lactaldehyde isomers with different results. The only three C$_3$H$_6$O$_2$ molecules found in the ISM were ethyl formate \citep{Belloche2009,Tercero2013}, methyl acetate \citep{Tercero2013}, and hydroxyacetone \citep{Zhou2020}.
Three carbon chain aldehydes discovered in the ISM are propanal \citep{Hollis2004,Goesmann2015,Lykke2017,Yarnall2020}, propenal \citep{Dickens2001,Hollis2004,Agundez2021, Manigand2021}, and propynal \citep{Irvine1988}. Propenal is presumed a prebiotic species not only because of its formation during the decomposition of sugars \citep{Moldoveanu2010,Bermudez2013}, but also because it acts as an intermediate in the prebiotic synthesis of the amino acid methionine \citep{VanTrump1972}. Propenal also contains a vinyl group, which is a structural motif present in several interstellar molecules; one of the examples is vinylalcohol \citep{Turner2001,Agundez2021}.

The (sub-/)millimetre-wave studies of aldehydes are not strictly limited to the astronomical applications but also serve for better understanding of the effects in the spectra, such as large amplitude motions or perturbations \citep{Daly2015,ZINGSHEIM2017,ZINGSHEIM2022}.

\begin{figure}[t]
   \centering
   \includegraphics[trim = 5mm 5mm 5mm 5mm, clip, width=8.5cm]{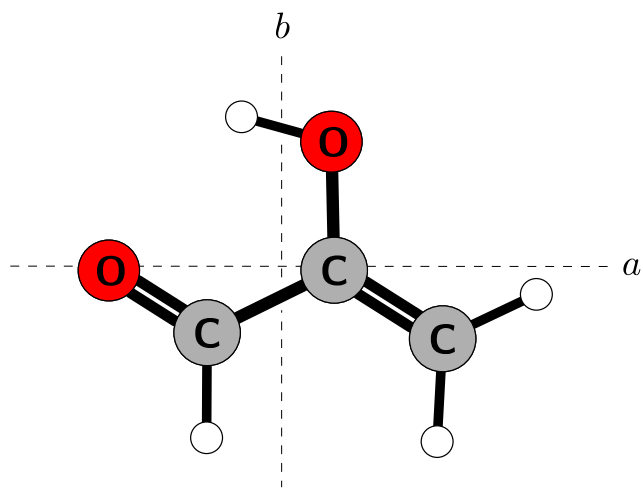}
      \caption{Structure of 2-hydroxyprop-2-enal displayed in $ab$ inertial plane.
      }
      \label{structure}
   \end{figure}

\begin{figure}[t]
   \centering
   \includegraphics[width=8.5cm]{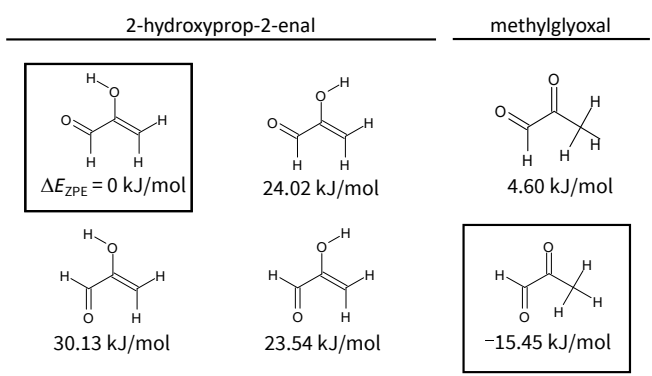}
      \caption{Keto-enol tautomeric relationship of 2-hydroxyprop-2-enal and methylglyoxal. The most stable isomers are highlighted. The structures are accompanied by their relative zero-point corrected energy (with respect to the most stable conformer of 2-hydroxyprop-2-enal) calculated at B3LYP/6-311++G(d,p) level of theory and basis set.
      }
      \label{isomers}
   \end{figure}

Our molecule of interest, 2-hydroxyprop-2-enal (or 2-hydroxyacrylaldehyde, see Fig. \ref{structure}), combines the structural motifs of the aforementioned vinyl alcohol and propenal. It belongs to the C$_3$H$_4$O$_2$ family of isomers. The isomerization enthalpies for several members of this family were calculated by \cite{KARTON2014}. Some stable C$_3$H$_4$O$_2$ isomers -- acrylic acid \citep{Calabrese2015, Alonso2015, Coutens2022}, vinyl formate \citep{Alonso_2016, Coutens2022} and 3-hydroxyprop-2-enal \citep{Coutens2022} -- have already been the subject of astrophysical investigation. Another attractive feature of 2-hydroxyprop-2-enal is its tautomeric relationship with methylglyoxal; see Fig. \ref{isomers}. Methylglyoxal was discovered, for example, in precometary ice analogues \citep{doi:10.1073/pnas.1418602112}. In addition, the prebiotic importance of 2-hydroxyprop-2-enal lies in methylglyoxal synthetase dependent decomposition of dihydroxyacetone phosphate, where it is created as a transient intermediate \citep{Rose2002}. In addition, it has been suggested that 2-hydroxyprop-2-enal is formed in irradiated methanol and methanol-carbon monoxide ice analogues \citep{Maity2015}. The microwave spectra of 2-hydroxyprop-2-enal were studied for the very first time in 2003 \citep{Lovas2003} up to 25 GHz. The species was identified as a decomposition product of glyceraldehyde at a nozzle temperature of 135~$^{\circ}$C.

The predictions of the rotational transitions to the frequency range of common surveys, such as surveys performed with the Atacama Large Millimeter/submillimeter Array (ALMA), like the
Re-Exploring Molecular Complexity with ALMA survey (ReMoCA) that covers the frequency range 84--114~GHz \citep{Belloche19} or the Protostellar Interferometric Line Survey (PILS) covering 329--363~GHz \citep{Jorgensen2016}, are rather unreliable if they are based on low-frequency data. Despite a slight setback in finding its parent molecules in space, there were other similar species with successful detection in the ISM. The data from the ALMA-PILS survey \citep{Coutens2022} indicate that 2-hydroxyprop-2-enal is a substantiate candidate molecule for interstellar search.

The outline of the paper is as follows: The experimental details on the preparation of the compound and measurements are described in Sect.~\ref{s:experiments}, notes about quantum-chemical calculations are to be found in Sect.~\ref{s:QChem}, the spectral analysis of the ground state and the lowest-lying vibrational state is provided in Sect.~\ref{s:analysis}, the searches for 2-hydroxyprop-2-enal in the interstellar medium are covered in Sects.~\ref{s:astro1} and ~\ref{s:astro2}, Sect.~\ref{s:conclusions} is dedicated to our conclusions.

\section{Experiments}
{\label{s:experiments}}
\subsection{Species preparation}
The compound 2-hydroxyprop-2-enal was prepared \textit{in situ} during the experiment. Initially, we have tried two ways to synthesise the species according to the results reported by \cite{Lovas2003}. Both are based on dehydration of the parent molecule, that is, glyceraldehyde (HOCH$_2$CH(OH)CHO) or dihydroxyacetone (HOCH$_2$C(O)CH$_2$OH). The precursors were purchased from Sigma Aldrich. Either of them proved to be a good source for producing the desired species in our experiment. Dihydroxyacetone as a precursor turned out to be a cost-effective solution, and therefore, all further measurements were carried out with it. The precursor in the solid phase was heated in a glass vessel to approximately 65~$^{\circ}$C, so that the rotational lines of 2-hydroxyprop-2-enal were distinguished and dihydroxyacetone did not melt. Apart from the aimed 2-hydroxyprop-2-enal, the spectrum was overflowing with, of course, the unreacted precursor and by-products, such as methanol, ethanol, or $trans$-methylglyoxal. $Trans$-methylglyoxal was also observed in the original microwave study \citep{Lovas2003}.
\subsection{Experiment}
The spectrum of 2-hydroxyprop-2-enal was recorded with the upgraded Prague semiconductor millimetre-wave spectrometer described elsewhere \citep{Kania2006} in two spectral ranges (128--166 GHz and 285--329 GHz). The sample pressure in the measuring glass cell varied from 15 to 30~$\mu$bar. Two different glass cells with lengths of 2.3 m and 2.8 m were used. The optical path of the cells was doubled to 4.6~m and 5.6~m, respectively, using a roof-top mirror. For the spectral measurements we adopted the same procedure like for 2-iminopropanenitrile \citep{Lukova2022} which also revealed relatively low stability in the cell. The sample was introduced to the chamber and kept there without pumping until the intensity of a selected rotational line decreased by 25~\% of its original intensity. Subsequently, the measuring cell was evacuated and filled again with the fresh sample. A frequency modulation of 28~kHz and a second harmonic lock-in detection were employed. The accuracy of the isolated measured rotational lines was 30~kHz.

\section{Quantum-chemical calculations}
{\label{s:QChem}}

Although the values of the rotational constants together with the quartic centrifugal distortion constants for 2-hydroxyprop-2-enal are known, we have performed quantum-chemical calculations employing the density functional theory (DFT) combined with the B3LYP functional and Pople’s basis set with added polarization and diffusion functions (6-311++G(d,p)). The structure optimization, analytical calculations of second energy derivatives followed by their numerical differentiation by finite-difference methods were performed using very tight convergence criteria and ultrafine grid.  The harmonic and anharmonic force field calculations up to cubic level led to estimations of, for example, vibrational energies and sextic centrifugal distortion constants that supported the spectral analysis. All calculations were carried out with Gaussian 16 \citep{g16}.
The four lowest-energy conformers of 2-hydroxyprop-2-enal and their spectroscopic properties are shown in Table~\ref{A0}.

\section{Spectroscopic results and discussion}
{\label{s:analysis}}

\subsection{Ground state}
{\label{ss:gs}}
\begin{figure*}[ht]
   \centering
   \includegraphics[trim = 5mm 5mm 5mm 5mm, clip, width=17.5cm]{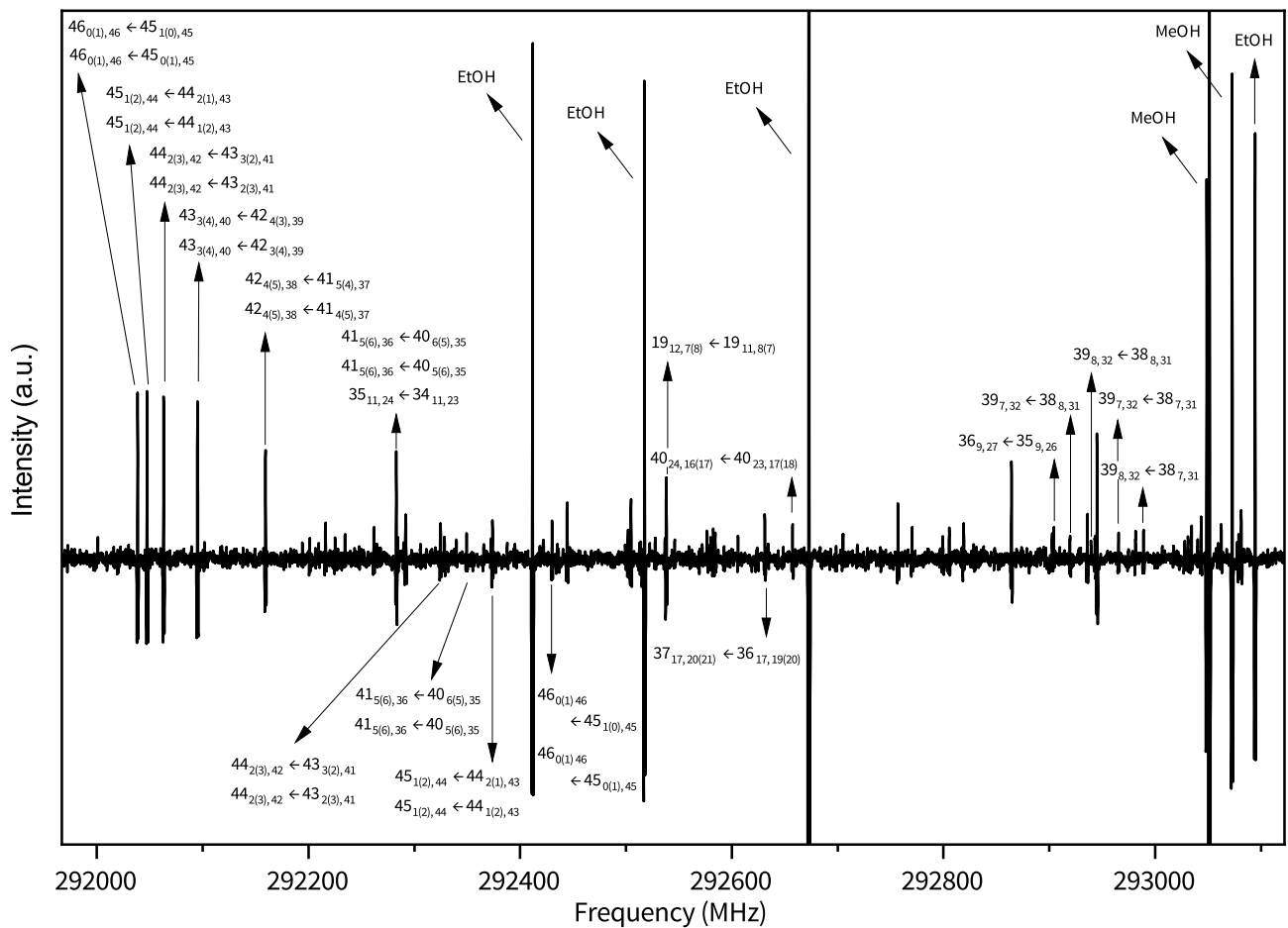}
      \caption{A segment of spectra showing an overview of coalesced and distinguished lines. Arrows pointing upwards belong to the ground state transitions, arrows pointing downwards represent excited vibrational state. Several transitions of ethanol (EtOH) and methanol (MeOH) are shown as well.
      }
      \label{2hp_overview}
   \end{figure*}

2-hydroxyprop-2-enal belongs to the $C_s$ symmetry point group with Ray’s asymmetry parameter of $\kappa =$ --0.6, making it a prolate asymmetric rotor. It is a planar molecule with heavy atoms located in the $ab$ inertial plane as shown in Fig.\ref{structure}. The previously determined values of electric dipole moments were $|\mu_{a}|=$~1.146(10)~D, $|\mu_{b}| =$~1.560(12)~D, and $|\mu_{c}| =$~0~D from symmetry \citep {Lovas2003}. Due to the high density of rotational lines originating not just from 2-hydroxyprop-2-enal, but also from the aforementioned molecules, we started the analysis in the 285 GHz region where easily identifiable groups of transition lines were expected to be observed. In series with low values of $K_a$ quantum number, each line comprises two $b$-type and two $a$-type transitions, as shown in Fig.\ref{2hp_overview}.

As this quantum number progressively increases, the lines are separated into quadruplets with higher intensity $b$-type lines surrounding $a$-type transitions. With the initial set of transitions being fitted and predictions made, we proceeded to analyse the lower frequency part of the spectrum where we sought resembling quadruplet patterns. Apart from the $R$-branch transitions, rotational lines belonging to $Q$-branch transitions were searched and analysed as well. The search was supported by graphical help of the Loomis-Wood type plot from Kisiel’s Assignment and Analysis of Broadband Spectra package \citep{Kisiel2005,Kisiel2012}.
Together with the data from the previous work \citep{Lovas2003}, a total set of 968 rotational transitions was employed in the final fit, out of which 594 belong to the $R$-branch with 340 $a$- and 254 $b$-type transitions, and 374 belong to the $Q$-branch $b$-type transitions. Lines in close proximity to the transitions of interfering species and lines with deformed shapes in both branches were excluded.
The Pickett’s program package SPFIT/SPCAT \citep{PICKETT1991} with Watson’s $A$-reduced Hamiltonian in I$^{\text{r}}$ representation were employed throughout the analysis. The resulting constants are summarised in Table~\ref{T1}.

The analysis of the ground state provides an extended set of spectroscopic constants, which is in an excellent agreement with quantum-chemical calculations as depicted in Table~\ref{T1}. All constants were improved by up to three orders of magnitude and a set of sextic centrifugal distortion constants is determined for the first time. 

\begin{table*}
\caption{Spectroscopic constants of 2-hydroxyprop-2-enal in the ground state (G.S.) and the lowest lying excited vibrational state ($A$-reduction, I$^{\text{r}}$-representation).}
\label{T1}
\begin{center}
\begin{footnotesize}
\setlength{\tabcolsep}{5.0pt}
\begin{tabular}{ l r r r r r }
\hline\hline
 & \multicolumn{3}{c} {G.S.} & &   \multicolumn{1}{c} {$v_{\text{21}}=1$}  \\
 \cline{2-4} \cline{6-6}
& \multicolumn{1}{c} {\cite{Lovas2003}} & \multicolumn{1}{c}{This work} & \multicolumn{1}{c}{Theory\tablefootmark{a}} && \multicolumn{1}{c}{This work} \\
 \hline
$A                   $  /               MHz     &  10\,201.6867(12)   &  10\,201.68699(22)\tablefootmark{b} & 10\,206.420 &&  10\,119.20549(88)                  \\
$B                   $  /               MHz     &   4543.3353(22)   &  4543.334833(81)                  & 4542.237  &&   4541.30508(18)              \\
$C                   $  /               MHz     &    3141.7866(20)  &  3141.787342(86)                  & 3143.335  &&   3146.37761(18)              \\
$\mathit{\Delta_{J}}          $  /               kHz     &     0.898(60)     & 0.898205(72)                      &    0.882  &&      0.91647(17)              \\
$\mathit{\Delta_{JK}}         $  /               kHz     &     5.99(40)      & 6.03919(20)                       &    6.138  &&      5.99556(97)               \\
$\mathit{\Delta_{K}}          $  /               kHz     &      5.08(25)     & 5.07237(91)                       &    4.737  &&      3.0074(33)                    \\
$\delta_{J}          $  /               kHz     &       0.268(15)   & 0.259966(21)                      &    0.254  &&      0.266214(22)                   \\
$\delta_{K}          $  /               kHz     &       4.44(94)    & 4.25939(31)                       &    4.215  &&      3.76131(84)                     \\
$\mathit{\Phi_{J}}            $  /               mHz     &                   & 0.122(19)                         &    0.115  &&      0.095(47)                     \\
$\mathit{\Phi_{JK}}           $  /               mHz     &                   & 2.46(11)                          &    2.146  &&    --11.15(41)                        \\
$\mathit{\Phi_{KJ}}           $  /               mHz     &                   &--64.22(29)                        & --62.553  &&    --34.2(33)                         \\
$\mathit{\Phi_{K}}            $  /               mHz     &                   & 105.35(93)                        &  103.965  &&     [105.35]\tablefoottext{c}          \\
$\phi_{J}            $  /               mHz     &                   &  0.0909(48)                       &    0.089  &&      [0.0909]\tablefoottext{c}       \\
$\phi_{JK}           $  /               mHz     &                   &  1.76(10)                         &    1.630  &&      3.40(34)        \\
$\phi_{K}            $  /               mHz     &                   & 49.18(43)                         &   46.829  &&     --48.4(25)         \\
$J_{\text{min}}/J_{\text{max}}$                 & 0 / 4             &  1 / 66                           &           &&    7 / 61              \\
$K_{a}^{\text{min}}/K_{a}^{\text{max}}$         & 0 / 1             & 0 / 33                            &           &&   0 / 20               \\
$N$\tablefootmark{d}                            & 15                & 669                               &           &&   295                 \\
$\sigma_{\text{fit}}$\tablefootmark{e}/ MHz     &                   & 0.024                             &           &&   0.032                \\
$\sigma_{\text{w}}$\tablefootmark{f}            & 0.58              & 0.70                              &           &&   0.87                 \\
\hline
\end{tabular}
\end{footnotesize}
\end{center}
\tablefoot{
\tablefoottext{a}{B3LYP/6-311++G(d,p).}
\tablefoottext{b}{The numbers in parentheses are 1$\sigma$ uncertainties (67\% confidence level) in units of the last decimal digits. The SPFIT/SPCAT program package was used for the analysis.}
\tablefoottext{c}{Fixed to the ground state value, which is usually a preferred constraint against the zero or poorly determined value \citep{Urban1990}.}
\tablefoottext{d}{Number of distinct frequency lines in the fit.}
\tablefoottext{e}{Root mean square deviation of the fit.}
\tablefoottext{f}{Unitless (weighted) deviation of the fit.}}
\end{table*}

The partition functions needed to calculate the column densities or their upper limits are provided in Table~\ref{T2}. The rotational partition functions for different temperatures were obtained from SPCAT covering all rotational states up to $J$~=~100. The vibrational partition functions were calculated using Eq. 3.60 from \cite{Gordy1970} employing the anharmonic frequencies of 21 vibrational modes shown in Table~\ref{A.1}.

\begin{table}
\caption{Partition functions for the ground state of 2-hydroxyprop-2-enal.}
\label{T2}
\begin{center}
\begin{footnotesize}
\setlength{\tabcolsep}{8.0pt}
\begin{tabular}{ r r r }
  \hline\hline
  \multicolumn{1}{c} {T (K)} & \multicolumn{1}{c} {$Q_\text{rot}$} & \multicolumn{1}{c} {$Q_\text{vib}$} \\ \hline
  300 & 72\,476.6361 & 3.52 \\
  225 & 47\,196.4989 & 2.03 \\
  150 & 25\,695.6466 & 1.32 \\
  75  & 9083.9691  & 1.03 \\
  37.5  & 3213.1188  & 1.00 \\
  18.75  & 1137.4106  & 1.00 \\
  9.375   & 403.1962   & 1.00 \\
  \hline
\end{tabular}
\end{footnotesize}
\end{center}
\end{table}

\subsection{Excited vibrational state $v_{\text{21}}=1$}

Our quantum-chemical calculations identified the lowest-frequency vibrational mode $\nu_{\text{21}}$ as the twisting skeletal vibration with the anharmonic frequency of 196.5~cm$^{-1}$. The first excited state of this mode was thus expected to be sufficiently populated at the temperature of the experiment. The predictions of its rotational constants relied on theoretical estimations of vibrational-rotational changes of the rotational constants with respect to the ground state from the first column of Table~\ref{T3}. Rotational transitions in $v_{\text{21}}=1$ were displaced to higher frequencies from the ground state and the Loomis-Wood-type diagrams greatly facilitated their assignments.
Only the most intense and well distinguished lines were analysed. The data set for $v_{\text{21}}=1$ consists of 487 rotational transitions up to $K_a=20$. The $R$-branch transitions are covered by 110 $b$-type transitions and 196 $a$-type transitions. The remaining 181 transitions belong to $b$-type $Q$-branch transitions. Not all of the visible series were added to the fit, as the signal-to-noise ratio was rather poor. The resulting spectroscopic constants are given in Table~\ref{T1}. The inertial defect $\mathit{\Delta}$~=~--0.611~u{\AA}$^2$ obtained from the rotational constants is more negative than its ground state counterpart ($\mathit{\Delta}$~=~0.083~u{\AA}$^2$) which confirms the out-of-plane character of $\nu_{\text{21}}$ vibrational mode.

\begin{table}
\caption{Differences between rotational constants of $v_{\text{21}}=1$ and the ground state of 2-hydroxyprop-2-enal.}
\label{T3}
\begin{center}
\begin{footnotesize}
\setlength{\tabcolsep}{5.0pt}
\begin{tabular}{ l r r }
\hline\hline
 & \multicolumn{1}{c} {Calculated \tablefoottext{a}} & \multicolumn{1}{c} {Experimental} \\
\hline
$A_{v}$ -- $A_{0}$       /               MHz       &  --84.54 &   --82.48     \\
$B_{v}$ -- $B_{0}$       /               MHz       &   --1.47 &   --2.27     \\
$C_{v}$ -- $C_{0}$       /               MHz       &   4.65   &   4.59  \\

\hline
\end{tabular}
\end{footnotesize}
\end{center}
\tablefoot{
\tablefoottext{a}{Calculated at B3LYP/6-311++G(d,p) level of theory.}
}
\end{table}

Most of the values for constants of the $v_{\text{21}}=1$ state are in agreement with those for the ground state presented in Table~\ref{T1}. Yet several discrepancies occur:
A difference from the values acquired by the ground state analysis is evident for the quartic centrifugal distortion constant $\Delta_{K}$ and for the $K$-dependent sextic constants. It is possible that the centrifugal distortion constants which are dependent on this quantum number display substantial variation with vibrational state
like in acrylonitrile \citep{Cazzoli1988,Kisiel2009,Kisiel2012}. However, the possible interaction with the nearby excited vibrational state at 277.4 cm$^{-1}$, or even a combination of both effects cannot be excluded either.
Unfortunately, the insufficient intensities for other low-lying vibrational states made their analysis impossible, thus hampering further assessment of this issue. Nevertheless, the Loomis-Wood type plots corroborate our assignment of the rotational lines.

The list of measured transitions for both ground state and $\nu_{\text{21}}=1$ state are listed in Table 5.

\section{Search for 2-hydroxyprop-2-enal towards IRAS16293-2422}
\label{s:astro1}

\subsection{Observations}
\label{ss:observations}
The Protostellar Interferometric Line Survey (PILS) is an unbiased spectral survey of the solar-type protostar IRAS16293-2422 carried out with ALMA between 329.1 and 362.9 GHz with a spectral resolution of $\sim$0.2 km\,s$^{-1}$ \citep{Jorgensen2016}. The high spatial resolution of $\sim$0.5$\arcsec$ clearly disentangles the emission of the two components A and B, which are separated by $\sim$5$\arcsec$. The data reduction process is detailed in \citet{Jorgensen2016}. Thanks to its high sensitivity (rms $\sim$ 4-5 mJy in a bin of 1 km s$^{-1}$), a large variety of species were detected for the first time in low-mass protostars through this survey (for example, \citealt{Lykke2017, Fayolle2017,Coutens2018,Coutens2019,Manigand2020}). In particular, several lines were recently assigned to 3-hydroxypropenal, HOCHCHCHO, towards component B \citep{Coutens2022}.
The chemical network presented in \citet{Manigand2021} was then updated to include 3-hydroxypropenal and several of its isomers including 2-hydroxyprop-2-enal \citep{Coutens2022}. The chemical model was in good agreement with the column density derived for 3-hydroxypropenal (within a factor 5) and the upper limits derived for vinyl formate (C$_2$H$_3$OCHO) and 2-propenoic acid (C$_2$H$_3$COOH). The spectroscopic data were missing for the other isomers. A high abundance of 2-hydroxyprop-2-enal (6.6\,$\times$\,10$^{-4}$ with respect to CH$_3$OH) was predicted by the model, which suggested that this species could be detectable in this source.

\subsection{Nondetection of 2-hydroxyprop-2-enal }
\label{ss:2-hydroxyprop-2-enalnondetection}

Thanks to the spectroscopic measurements obtained in Sect.\ref{s:analysis}, we consequently searched for 2-hydroxyprop-2-enal towards IRAS16293 B at the exact same position where the lines of 3-hydroxypropenal were identified ($\alpha_{\rm J2000}$ = 16$^{\rm h}$32$^{\rm m}$22$\fs$58, $\delta_{\rm J2000}$ = -24$\degr$28$\arcmin$32.8$\arcsec$). No clear detection is obtained for this species. We used for that the CASSIS\footnote{CASSIS has been developed by IRAP-UPS/CNRS \citep{Vastel2015}. \url{http://www.cassis.irap.omp.eu} } software. We assumed a FWHM of 1 km\,s$^{-1}$ and a source size of 0.5$\arcsec$ to determine the upper limit. Complex organic molecules in this object show excitation temperatures, $T_{\rm ex}$, between 125 and 300\,K \citep{Jorgensen2018}. While some lines could correspond to 2-hydroxyprop-2-enal for $T_{\rm ex}$= 300 K, others are missing with the same model. The 3$\sigma$ upper limit is $\sim$2.4\,$\times$\,10$^{15}$ cm$^{-2}$ based on the undetected lines and taking into account the vibrational correction (see Table~\ref{T2}). For 125 K, which was slightly favoured for 3-hydroxypropenal \citep{Coutens2022}, the lines of 2-hydroxyprop-2-enal are usually blended with other species. The upper limit is, in this case, $\sim$9.4\,$\times$\,10$^{14}$ cm$^{-2}$. The upper limit on its abundance with respect to CH$_3$OH is consequently about $\sim$2.4\,$\times$\,10$^{-4}$, that is a factor of about 3 lower than the predictions of the chemical model, but still within the uncertainties.
According to the model, 2-hydroxyprop-2-enal is mostly formed by the tautomerisation of methyl glyoxal (CH$_3$COCHO) itself produced essentially by the reaction between CH$_3$CO and HCO. The association of HCO and CH$_2$COH on grains, leading directly to 2-hydroxyprop-2-enal, is negligible because CH$_2$COH has a very low concentration. The derived abundance is still within the uncertainties of the chemical model which is estimated to be a factor 20 for this species. Indeed, the uncertainty of the branching ratios of the association reactions is estimated to be a factor 5, while an uncertainty of a factor 4 comes from the destruction reactions with hydrogen atoms. As explained in \citet{Coutens2022}, the tautomerisation of methyl glyoxal could be less favourable on grains than assumed in the model because this tautomerisation involves a transition state quite high in energy. The uncertainty on the production of 2-hydroxyprop-2-enal is much larger than the uncertainty on the production of methyl glyoxal (CH$_3$COCHO). The upper limit derived for the 2-hydroxyprop-2-enal/3-hydroxypropenal abundance ratio ($\lesssim$ 0.9 for $T_{\rm ex}$= 125 K and $\lesssim$ 1.3 for $T_{\rm ex}$= 300 K) is consistent with the value predicted by the chemical model ($\sim$ 1.4), taking into account the uncertainties described above.

\section{Search for 2-hydroxyprop-2-enal and related aldehydes toward Sgr~B2(N1)}
\label{s:astro2}

\subsection{Observations}
\label{ss:obs_remoca}

We used astronomical data obtained with ALMA toward the high-mass star forming
protocluster Sgr~B2(N). This data set constitutes the imaging spectral line
survey ReMoCA. A detailed description of
the observations and data reduction was reported in \citet{Belloche19}. The
main characteristics of the survey are the following. The phase center is
located at the equatorial position ($\alpha, \delta$)$_{\rm J2000}$=
($17^{\rm h}47^{\rm m}19{\fs}87, -28^\circ22'16{\farcs}0$) which is half-way
between the two hot molecular cores Sgr~B2(N1) and Sgr~B2(N2). The survey was
split into five independent frequency setups to cover the frequency range from
84.1~GHz to 114.4~GHz with a spectral resolution of 488~kHz (1.7 to
1.3~km~s$^{-1}$). The median sensitivity per spectral channel achieved for this
survey is 0.8~mJy~beam$^{-1}$, with a median angular resolution
(HPBW) of 0.6$\arcsec$ that corresponds to $\sim$4900~au at the distance
of Sgr~B2 \citep[8.2~kpc,][]{Reid19}. We used a slightly improved version of
the data reduction as described in \citet{Melosso20}.

We followed the strategy of \citet{Belloche19} and analysed the
spectrum obtained toward the position Sgr~B2(N1S) at
($\alpha, \delta$)$_{\rm J2000}$=
($17^{\rm h}47^{\rm m}19{\fs}870$, $-28^\circ22\arcmin19{\farcs}48$). This
position is offset by about
1$\arcsec$ to the south of the main hot core Sgr~B2(N1). It has a lower
continuum opacity compared to the peak of the hot core which facilitates the
detection of spectral lines of molecules. We assumed local
thermodynamic equilibrium (LTE) and produced synthetic spectra with the
astronomical software Weeds \citep[][]{Maret11} in order to analyse the
observed spectrum. The high densities of the regions where hot-core emission
is detected in Sgr~B2(N) \citep[$>1 \times 10^{7}$~cm$^{-3}$, see][]{Bonfand19}
justify our assumption of LTE.
We derived a best-fit synthetic spectrum for each molecule separately, and
then added the contributions of all identified molecules together. We used a
set of five parameters to model the contribution of each species: size of the
emitting region
($\theta_{\rm s}$), column density ($N$), temperature ($T_{\rm rot}$), linewidth
($\Delta V$), and velocity offset ($V_{\rm off}$) with respect to the assumed
systemic velocity of the source, $V_{\rm sys}=62$~km~s$^{-1}$.

\input{tab_ch2cohcho_weedsmodel.tex}

\begin{figure*}
\centerline{\resizebox{0.82\hsize}{!}{\includegraphics[angle=0]{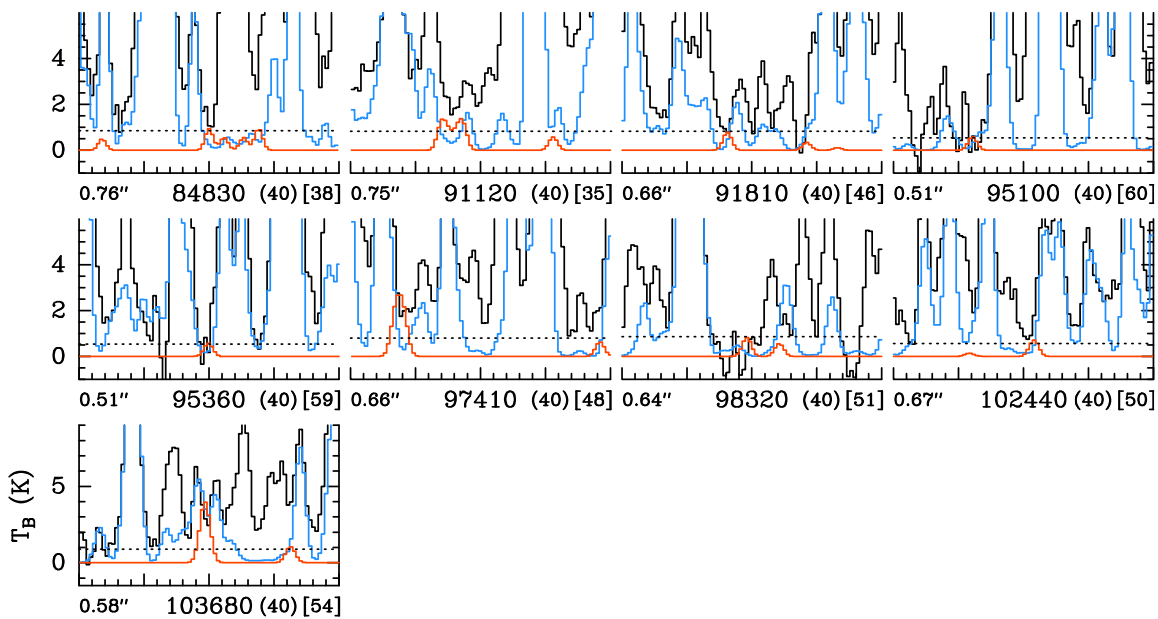}}}
\caption{Selection of transitions of CH$_2$C(OH)CHO, $\varv = 0$ covered
by the ReMoCA survey. The LTE synthetic spectrum used to derive the upper limit
on the column density of CH$_2$C(OH)CHO, $\varv = 0$ is
displayed in red and overlaid on the observed spectrum of Sgr~B2(N1S) shown in
black. The blue synthetic spectrum contains the contributions of all molecules
identified in our survey so far, but does not include the contribution of the
species shown in red. The central frequency is indicated in MHz below each
panel as well as the half-power beam width on the left, the width of each
panel in MHz in parentheses, and the continuum level of the
baseline-subtracted spectra in K in brackets. The y-axis is labelled in
brightness temperature units (K). The dotted line indicates the $3\sigma$
noise level.}
\label{f:spec_ch2cohcho_ve0}
\end{figure*}

\begin{figure}
\centerline{\resizebox{0.88\hsize}{!}{\includegraphics[angle=0]{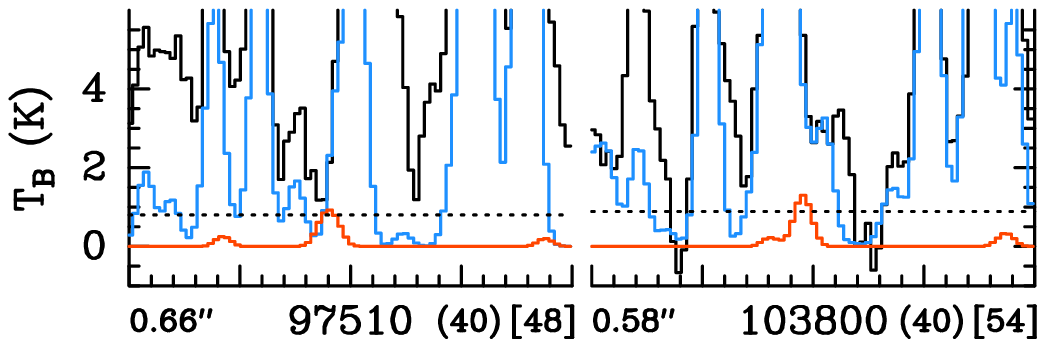}}}
\caption{Same as Fig.~\ref{f:spec_ch2cohcho_ve0}, but for
CH$_2$C(OH)CHO, $\varv_{21} = 1$.}
\label{f:spec_ch2cohcho_v21e1}
\end{figure}

\subsection{Nondetection of 2-hydroxyprop-2-enal}
\label{ss:nondetection_remoca}

In order to search for 2-hydroxyprop-2-enal, CH$_2$C(OH)CHO, toward
Sgr~B2(N1S), we relied on the LTE parameters derived for acetaldehyde,
CH$_3$CHO, toward the same source from the ReMoCA survey, as reported in
\citet{SanzNovo22}. These parameters are listed in Table~\ref{t:coldens}.
Assuming that the more complex
molecule 2-hydroxyprop-2-enal traces the same region as acetaldehyde, we
produced LTE synthetic spectra for the former species adopting the same
parameters as for the latter with only the column density left as a free
parameter. We employed the frequency predictions derived for
2-hydroxyprop-2-enal using the spectroscopic constants from Table~\ref{T1} to compute its LTE synthetic spectrum.
The molecule is not detected toward Sgr~B2(N1S), as illustrated in
Figs.~\ref{f:spec_ch2cohcho_ve0} and \ref{f:spec_ch2cohcho_v21e1}. The upper limit on the total column density
of 2-hydroxyprop-2-enal obtained from the ReMoCA survey is reported in
Table~\ref{t:coldens}, after accounting for the vibrational partition function
that was computed with the information provided in Sect.~\ref{ss:gs}

We also report in Table~\ref{t:coldens} the column density upper limits that we
obtained with the ReMoCA survey for two related aldehydes: the more saturated
molecule 2-hydroxypropanal, CH$_3$CH(OH)CHO and 3-hydroxypropenal, HOCHCHCHO,
which is a structural isomer of 2-hydroxyprop-2-enal. None of these two
aldehydes is detected toward Sgr~B2(N1S), as illustrated in
Figs.~\ref{f:spec_ch3chohcho_ve0} and \ref{f:spec_hochchcho_ve0}, respectively.
We employed the spectroscopic entry 74519 (version 1) of the Cologne Database
for Molecular Spectroscopy\footnote{https://cdms.astro.uni-koeln.de/}
\citep[CDMS,][]{Mueller05} to compute the LTE synthetic spectra used to derive
the upper limit to the column density of 2-hydroxypropanal. This CDMS entry is
based on the measurements reported in \citet{Alonso19}. The upper limit reported
in Table~\ref{t:coldens} accounts for the (substantial) vibrational correction
that was estimated using the anharmonic frequencies of its normal vibrational modes
computed at the same level of theory and basis set (MP2/6-311++G(d,p)) as \citet{Alonso19} employed for
the harmonic ones.

The upper limit reported for 3-hydroxypropenal was obtained
using the CDMS entry 72504 (version 1) which relies on previous microwave and
submillimetre-wave measurements \citep{Baughcum1978,Stolze1983,Baba1999}.

The vibrational correction was estimated using the experimental
vibrational frequencies assigned by \citet{Tayyari98} and measured
by \citet{Smith83}. We also recall in
Table~\ref{t:coldens} the upper limit obtained for propanal toward Sgr~B2(N1S)
from the ReMoCA survey, as reported by \citet{SanzNovo22}.

Table~\ref{t:coldens} shows that 2-hydroxyprop-2-enal is at least 9 times less
abundant than acetaldehyde toward Sgr~B2(N1S). A similar limit is obtained for
its structural isomer 3-hydroxypropenal, while the limits obtained for
propanal and 2-hydroxypropanal are less stringent by nearly a factor of two. For comparison, the results reported in Sect. \ref{ss:2-hydroxyprop-2-enalnondetection} for 2-hydroxyprop-2-enal and those for 3-hydroxypropenal \citep{Coutens2022}, and acetaldehyde \citep{Jorgensen2018} indicate that the former two molecules are $>$130 and $\sim$120 times less abundant than acetaldehyde in IRAS16293 B, respectively. The limits obtained for both molecules in Sgr
B2(N1S) are thus less constraining by more than one order of magnitude compared to IRAS16293 B.

\section{Conclusions}
\textbf{\label{s:conclusions}}

We present the extended dataset for the ground state of 2-hydroxyprop-2-enal, together with the first data for the low-lying excited vibrational state $v_{\text{21}}=1$. The species was prepared \textit{in situ} by thermal decomposition of DHA and then characterised by millimetre-wave rotational spectroscopy. The effective Hamiltonian with quartic and sextic centrifugal distortion terms was employed within the analysis. The determination of sextic terms and the analysis of $\nu_{\text{21}}=1$ state were supported by quantum-chemical calculations. A total number of 669 and 295 distinct frequency lines covering 968 and 487 rotational transitions was used in the assignment of the ground and excited vibrational states, respectively.
These new laboratory measurements provided accurate frequency predictions to search for the molecule toward IRAS16293 B and Sgr~B2(N).
Additionally, emission lines of the related molecules 2-hydroxypropanal and 3-hydroxypropenal were searched for in the latter source.
The observational results are following:
\begin{enumerate}
\item We report a nondetection of 2-hydroxyprop-2-enal toward the protostar IRAS16293 B. The derived upper limit is in agreement within the uncertainties with the predictions of the chemical model developed for 3-hydroxypropenal in \citep{Coutens2022}. The 2-hydroxyprop-2-enal/3-hydroxypropenal abundance ratio is estimated to be $\lesssim$ 0.9-1.3 in agreement with the predicted value  of $\sim$ 1.4.
\item We report a nondetection of 2-hydroxyprop-2-enal toward
the main hot core of Sgr~B2(N) that was targeted with ALMA. We find that this
molecule is at least 9 times less abundant than acetaldehyde toward this
source. Its column density upper limit is nearly a factor of two lower than the
one derived earlier for propanal in the same source.
\item None of the related aldehydes 2-hydroxypropanal and 3-hydroxypropenal
was detected toward Sgr B2(N)'s main hot core. They are at least 4 and 10 times less
abundant than acetaldehyde, respectively.
\end{enumerate}
Despite the nondetections of 2-hydroxyprop-2-enal, the present work represents a significant improvement on previous laboratory studies below 25 GHz and provides
an excellent base for further future searches of this species in space.

\begin{acknowledgements}

The spectroscopic part of the article has been supported by the Czech Science Foundation (GACR, grant 19-25116Y). L.K, J.K., K.L, and K.V. gratefully acknowledge this funding. Computational resources were supplied by the project "e-Infrastruktura CZ" (e-INFRA CZ LM2018140 ) supported by the Ministry of Education, Youth and Sports of the Czech Republic.
This paper makes use of the following ALMA data:
ADS/JAO.ALMA\#2016.1.00074.S and ADS/JAO.ALMA\#2013.1.00278.S.
ALMA is a partnership of ESO (representing its member states), NSF (USA), and
NINS (Japan), together with NRC (Canada), NSC and ASIAA (Taiwan), and KASI
(Republic of Korea), in cooperation with the Republic of Chile. The Joint ALMA
Observatory is operated by ESO, AUI/NRAO, and NAOJ. The interferometric data
are available in the ALMA archive at https://almascience.eso.org/aq/.
Part of this work has been carried out within the Collaborative
Research Centre 956, sub-project B3, funded by the Deutsche
Forschungsgemeinschaft (DFG) -- project ID 184018867.
A.C. received financial support from the European Research Council (ERC) under the European Union’s Horizon 2020 research and innovation programme (ERC Starting Grant “Chemtrip”, grant agreement No 949278).

\end{acknowledgements}



\bibliography{2hp_aa}


\begin{appendix}
\label{appendix}

\section{Supplementary Tables}
\label{a:supplementarytables}
Table~\ref{A0} summarises the spectroscopic properties of the four lowest-energy conformers of 2-hydroxyprop-2-enal. Table~\ref{A.1} lists the frequencies of the normal vibrational modes of 2-hydroxyprop-2-enal.

\begin{table*}[h]
\caption{Four lowest-energy conformers of 2-hydroxyprop-2-enal.}
\label{A0}
\begin{center}
\begin{footnotesize}
\setlength{\tabcolsep}{5.0pt}
\begin{tabular}{llcccc}
\hline\hline
& & \multicolumn{4}{c} {Conformer} \\ \cline{3-6}
& &   I     & II       & III    & IV       \\
Constant  & Unit     & \includegraphics[width=1.7cm]{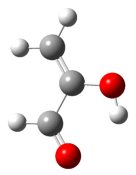}      & \includegraphics[width=1.7cm]{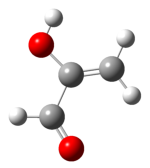}         &  \includegraphics[width=1.7cm]{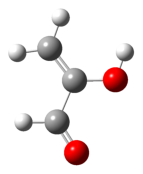}      &    \includegraphics[width=1.7cm]{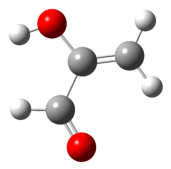}     \\
\hline
$A$ & MHz              & 10206 & 10038    & 9962   & 10044   \\
$B$ & MHz              & 4542  & 4264     & 4420   & 4239    \\
$C$ & MHz              & 3143  & 2993     & 3062   & 2981    \\
$|\mu_a|$ &  D         & 1.3   & 3.33     & 4.11   & 1.3     \\
$|\mu_b|$ &  D         & 1.65  & 0.51     & 1.29   & 1.4     \\
$|\mu_c|$ &  D         & 0     & 0        & 0      & 0       \\
$\Delta E_\text{ZPE}$ & kJ/mol & 0     & 23.54    & 24.02  & 30.13   \\
\hline
\end{tabular}
\end{footnotesize}
\end{center}
\end{table*}

\begin{table*}
\caption{List of harmonic and anharmonic frequencies for vibrational modes of
2-hydroxyprop-2-enal calculated by B3LYP/6-311++G(d,p).}
\label{A.1}
\begin{center}
\begin{footnotesize}
\setlength{\tabcolsep}{5.0pt}
\begin{tabular}{r r r c}
  \hline \hline
  & \multicolumn{2}{c} {Frequency (cm$^{-1}$)} & \\
  \cline{2-3}
  Mode & Harmonic & Anharmonic & Symmetry \\
  \hline
  1 & 3680.764 & 3490.279 & $A^{'}$ \\
  2 & 3257.119 & 3119.542 & $A^{'}$ \\
  3 & 3163.194 & 3017.537 & $A^{'}$ \\
  4 & 2974.271 & 2814.370 & $A^{'}$ \\
  5 & 1745.388 & 1712.650 & $A^{'}$ \\
  6 & 1707.794 & 1664.629 & $A^{'}$ \\
  7 & 1446.509 & 1406.382 & $A^{'}$ \\
  8 & 1397.543 & 1357.208 & $A^{'}$ \\
  9 & 1366.622 & 1340.344 & $A^{'}$ \\
 10 & 1255.587 & 1237.762 & $A^{'}$ \\
 11 &  981.869 &  964.669 & $A^{'}$ \\
 12 &  895.680 &  875.146 & $A^{'}$ \\
 13 &  681.221 &  665.809 & $A^{'}$ \\
 14 &  410.232 &  411.260 & $A^{'}$ \\
 15 &  282.018 &  277.358 & $A^{'}$ \\
 16 & 1000.470 &  990.702 & $A^{''}$ \\
 17 &  891.797 &  881.615 & $A^{''}$ \\
 18 &  732.165 &  716.236 & $A^{''}$ \\
 19 &  558.336 &  571.796 & $A^{''}$ \\
 20 &  524.626 &  471.809 & $A^{''}$ \\
 21 &  201.300 &  196.519 & $A^{''}$ \\
  \hline
\end{tabular}
\end{footnotesize}
\end{center}
\end{table*}
\clearpage

\section{Complementary figures: Nondetection of 2-hydroxypropanal and 3-hydroxypropenal towards Sgr B2(N1S)}
\label{a:spectra_n1s}

\begin{figure*}
\centerline{\resizebox{0.82\hsize}{!}{\includegraphics[angle=0]{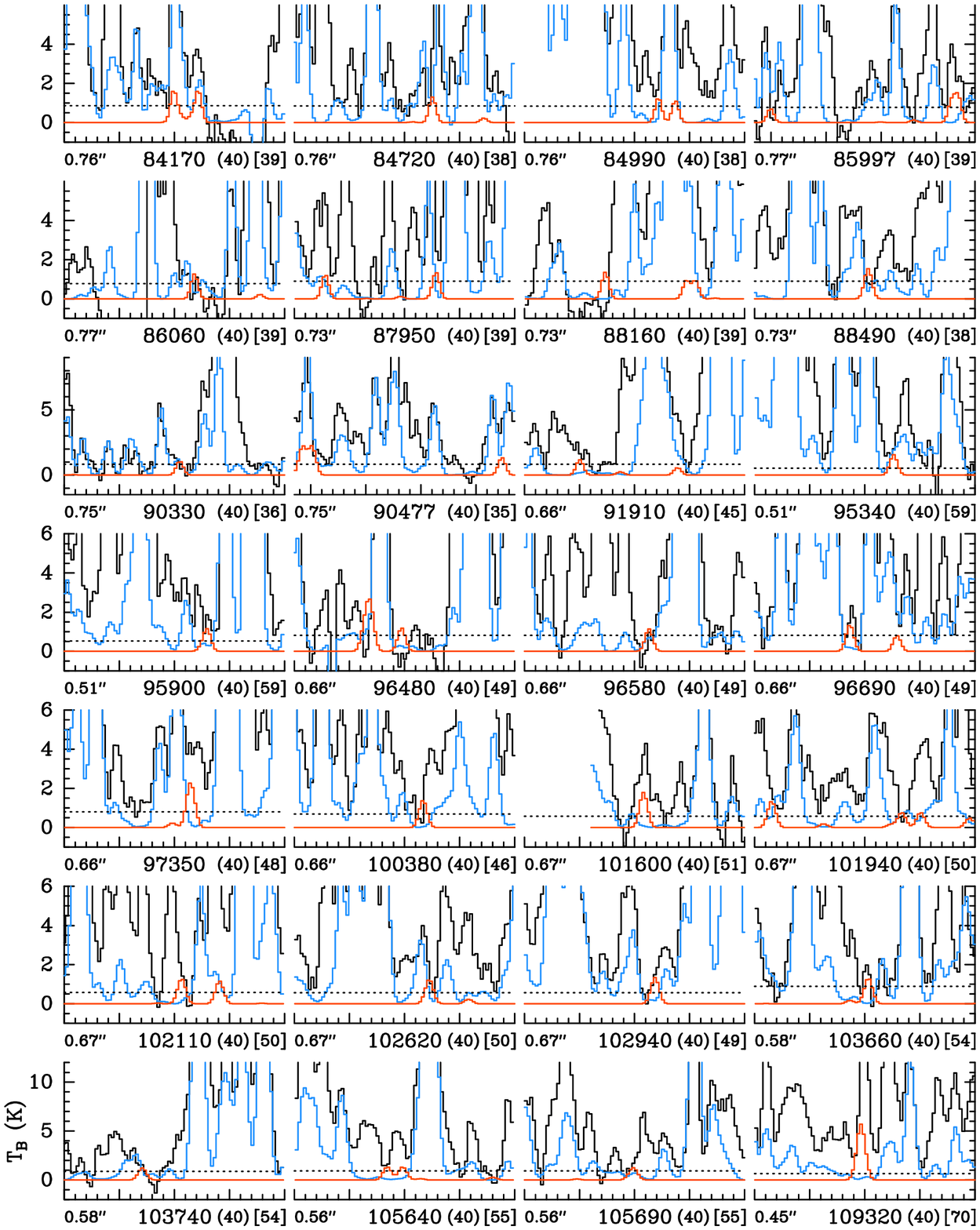}}}
\caption{Same as Fig.~\ref{f:spec_ch2cohcho_ve0}, but for CH$_3$CH(OH)CHO,
$\varv = 0$.
}
\label{f:spec_ch3chohcho_ve0}
\end{figure*}

Figures~\ref{f:spec_ch3chohcho_ve0} and \ref{f:spec_hochchcho_ve0} illustrate
the nondetection of 2-hydroxypropanal and 3-hydroxypropenal, respectively,
toward Sgr~B2(N1S) with the ReMoCA survey.

\begin{figure*}
\centerline{\resizebox{0.82\hsize}{!}{\includegraphics[angle=0]{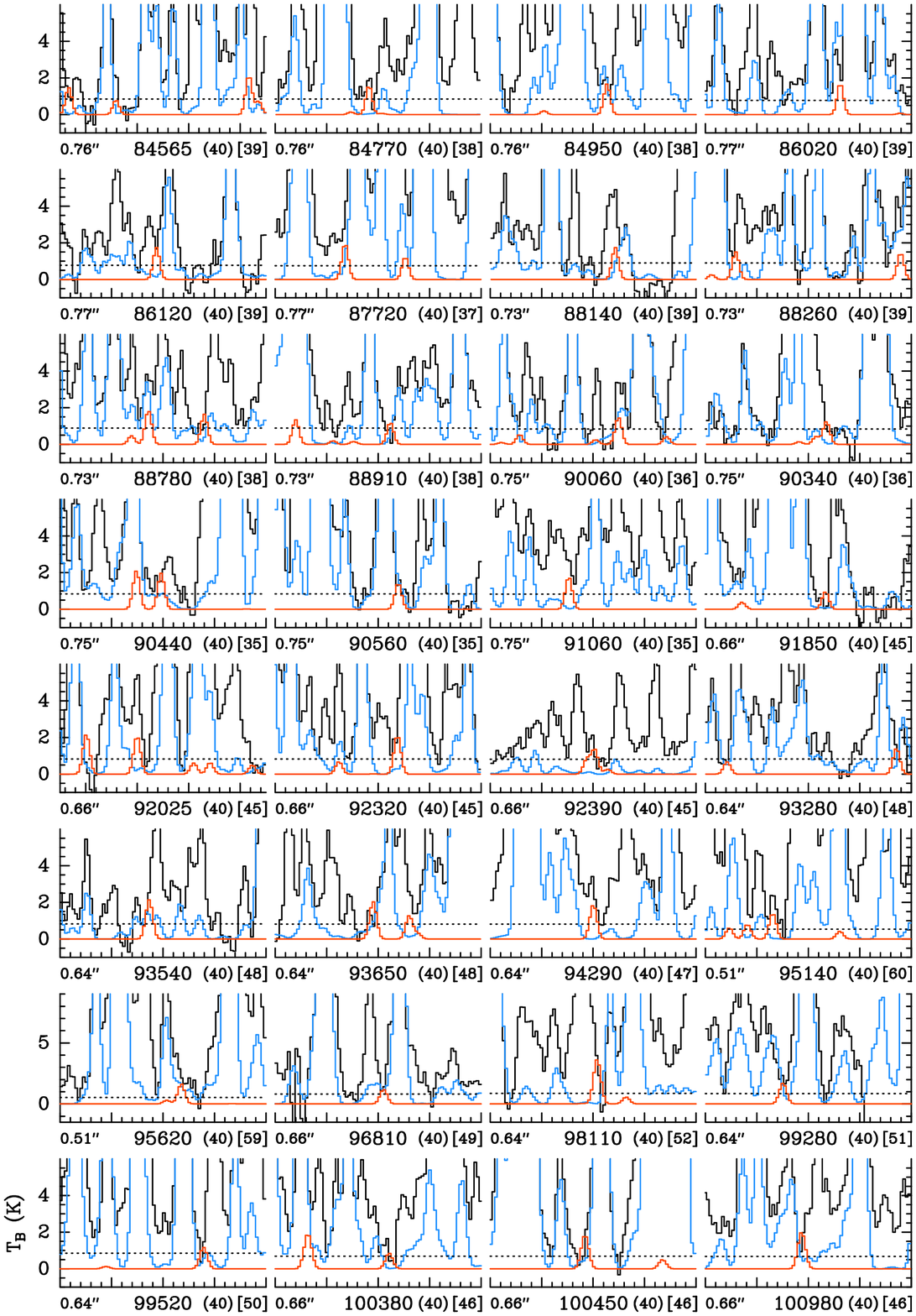}}}
\caption{Same as Fig.~\ref{f:spec_ch2cohcho_ve0}, but for HOCHCHCHO, $\varv$=0.}
\label{f:spec_hochchcho_ve0}
\end{figure*}

\begin{figure*}
\addtocounter{figure}{-1}
\centerline{\resizebox{0.82\hsize}{!}{\includegraphics[angle=0]{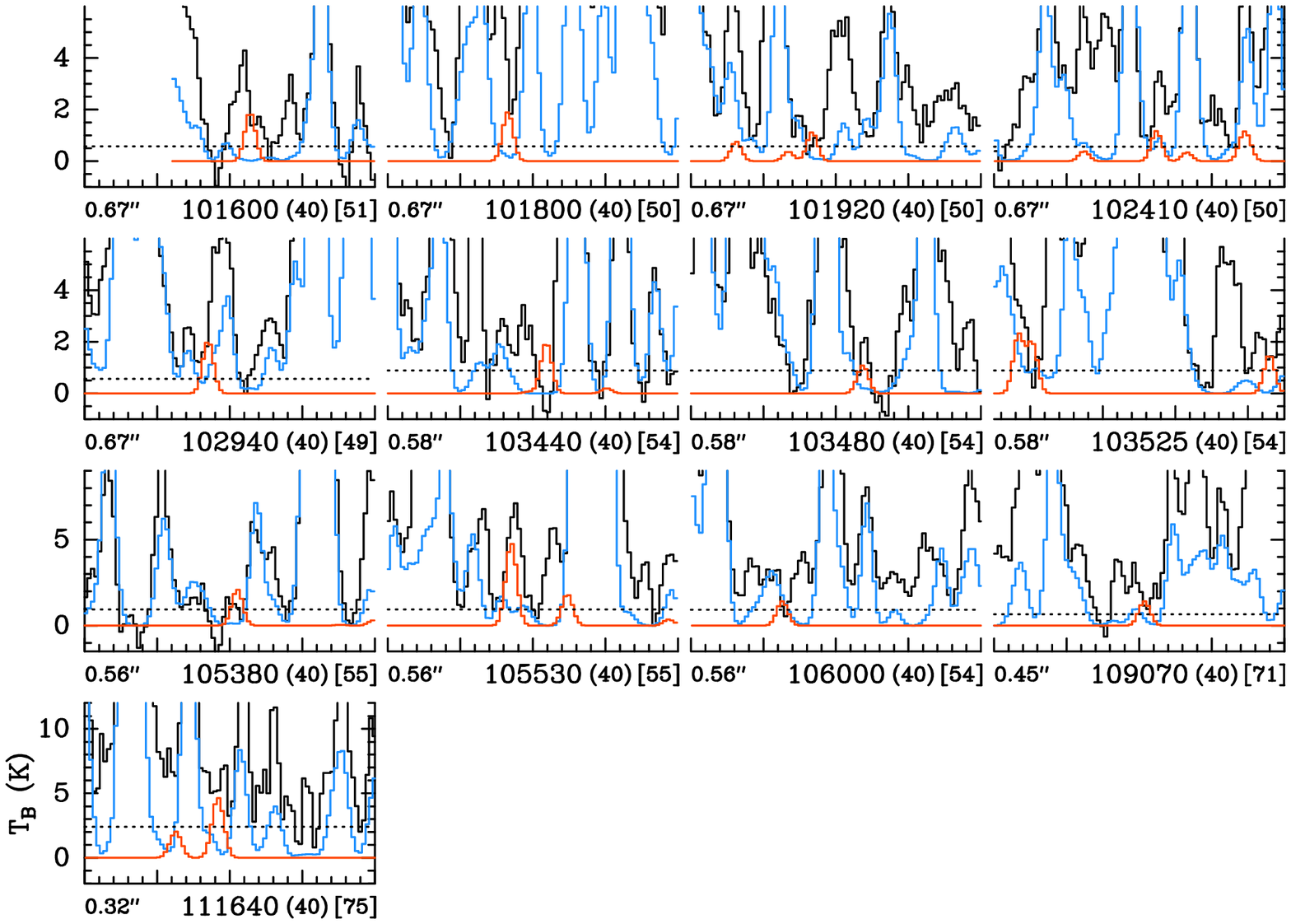}}}
\caption{continued.
}
\end{figure*}

\end{appendix}

\end{document}

%% file: tab_ch2cohcho_weedsmodel.tex
\begin{table*}[!ht]
 \begin{center}
 \caption{
 Parameters of our best-fit LTE model of acetaldehyde toward Sgr~B2(N1S) and upper limits for propanal, 2-hydroxypropanal, 2-hydroxyprop-2-enal, and 3-hydroxypropenal.
}
 \label{t:coldens}
 \vspace*{-1.2ex}
 \begin{tabular}{lcrccccccr}
 \hline\hline
 \multicolumn{1}{c}{Molecule} & \multicolumn{1}{c}{Status\tablefootmark{a}} & \multicolumn{1}{c}{$N_{\rm det}$\tablefootmark{b}} & \multicolumn{1}{c}{Size\tablefootmark{c}} & \multicolumn{1}{c}{$T_{\mathrm{rot}}$\tablefootmark{d}} & \multicolumn{1}{c}{$N$\tablefootmark{e}} & \multicolumn{1}{c}{$F_{\rm vib}$\tablefootmark{f}} & \multicolumn{1}{c}{$\Delta V$\tablefootmark{g}} & \multicolumn{1}{c}{$V_{\mathrm{off}}$\tablefootmark{h}} & \multicolumn{1}{c}{$\frac{N_{\rm ref}}{N}$\tablefootmark{i}} \\ 
  & & & \multicolumn{1}{c}{\small ($''$)} & \multicolumn{1}{c}{\small (K)} & \multicolumn{1}{c}{\small (cm$^{-2}$)} & & \multicolumn{1}{c}{\small (km~s$^{-1}$)} & \multicolumn{1}{c}{\small (km~s$^{-1}$)} & \\ 
 \hline
 CH$_3$CHO\tablefootmark{j}$^\star$ & d & 31 &  2.0 &  250 &  6.7 (17) & 1.09 & 5.0 & 0.0 &       1 \\ 
\hline 
 C$_2$H$_5$CHO\tablefootmark{j} & n & 0 &  2.0 &  250 & $<$  1.3 (17) & 4.46 & 5.0 & 0.0 & $>$     5.0 \\ 
\hline 
 CH$_3$CH(OH)CHO, $\varv=0$ & n & 0 &  2.0 &  250 & $<$  1.7 (17) & 5.68 & 5.0 & 0.0 & $>$     4.0 \\ 
\hline 
 CH$_2$C(OH)CHO, $\varv=0$ & n & 0 &  2.0 &  250 & $<$  7.2 (16) & 2.41 & 5.0 & 0.0 & $>$     9.3 \\ 
\hline 
 HOCHCHCHO, $\varv=0$ & n & 0 &  2.0 &  250 & $<$  7.0 (16) & 2.00 & 5.0 & 0.0 & $>$     9.6 \\ 
\hline 
 \end{tabular}
 \end{center}
 \vspace*{-2.5ex}
 \tablefoot{
 \tablefoottext{a}{d: detection, n: nondetection.}
 \tablefoottext{b}{Number of detected lines \citep[conservative estimate, see Sect.~3 of][]{Belloche16}. One line of a given species may mean a group of transitions of that species that are blended together.}
 \tablefoottext{c}{Source diameter (FWHM).}
 \tablefoottext{d}{Rotational temperature.}
 \tablefoottext{e}{Total column density of the molecule. $x$ ($y$) means $x \times 10^y$.}
 \tablefoottext{f}{Correction factor that was applied to the column density to account for the contribution of vibrationally excited states, in the cases where this contribution was not included in the partition function of the spectroscopic predictions.}
 \tablefoottext{g}{Linewidth (FWHM).}
 \tablefoottext{h}{Velocity offset with respect to the assumed systemic velocity of Sgr~B2(N1S), $V_{\mathrm{sys}} = 62$ km~s$^{-1}$.}
 \tablefoottext{i}{Column density ratio, with $N_{\rm ref}$ the column density of the previous reference species marked with a $\star$.}
 \tablefoottext{j}{The parameters were derived from the ReMoCA survey by \citet{SanzNovo22}.}
 }
 \end{table*}